\documentclass{emulateapj}

\newcommand{\rsoph}{RS Oph\space}

\shorttitle{X-ray extended emission from RS Oph}
\shortauthors{Luna et al.}

\begin{document}


\title{Chandra detection of extended X-ray emission \\
from the recurrent nova \object{RS Ophiuchi}}


\author{G. J. M. Luna\altaffilmark{1}, R. Montez\altaffilmark{2},
  J. L. Sokoloski\altaffilmark{3}, K. Mukai\altaffilmark{4} and
  J. H. Kastner\altaffilmark{2}}  






\altaffiltext{1}{Harvard-Smithsonian Center for Astrophysics, 60
  Garden St. MS 15, Cambridge, MA, 02138,
  USA. \email{gluna@cfa.harvard.edu}} 
\altaffiltext{2}{2100 Carlson Center for Imaging Science Rochester
  Institute of Technology Rochester, NY 14623 USA.} 
\altaffiltext{3}{Columbia Astrophysics Laboratory, 550 W. 220th
  Street, 1027 Pupin Hall, Columbia University, New York, NY 10027} 
\altaffiltext{4}{CRESST and X-ray Astrophysics Laboratory NASA/GSFC,
  Greenbelt, MD 20771, USA} 


\begin{abstract}

Radio, infrared, and optical observations of the 2006 eruption of the
symbiotic recurrent nova RS Ophiuchi (RS Oph) showed that the
explosion produced non-spherical ejecta. Some of this ejected material
was in the form of bipolar jets to the east and west of the central
source.  Here we describe X-ray observations taken with the $Chandra$
X-ray Observatory one and a half years after the beginning of the
outburst that reveal narrow, extended structure with a position angle
of approximately 300 degrees (east of north).
Although the orientation of the extended feature in the X-ray image is
consistent with the readout direction of the CCD detector, extensive
testing suggests that the feature is not an artifact. Assuming it is
not an instrumental effect, the extended X-ray structure shows hot
plasma stretching more than 1,900 AU from the central binary (taking a
distance of 1.6 kpc).
The X-ray emission is elongated in the northwest direction --- in line
with the extended infrared emission and some minor features in the
published radio image. It is less consistent with the orientation of
the radio jets and the main bipolar optical structure.
Most of the photons in the extended X-ray structure have energies of
less than 0.8 keV. If the extended X-ray feature was produced when the
nova explosion occurred, then its 1$^{\prime\prime}$.2 length as of
2007 August implies that
it expanded at an average rate of more than 2 mas d$^{-1}$, which
corresponds to a flow speed of greater than 6,000 km/s (d/1.6 kpc) in
the plane of the sky. This expansion rate is similar to the earliest
measured expansion rates for the radio jets.

\end{abstract}





\section{Introduction}

Novae are the most common major stellar explosions in the universe.
The basic properties of these explosions are well explained by models
in which a thermonuclear runaway at the base of the envelope of an
accreting white dwarf (WD) causes the WD to brighten to near the
Eddington luminosity and eject its envelope in a spherical outflow. An
increasing number of observations, however, are indicating that the
ejecta from many novae are anything but spherical
\citep[e.g.,][]{slavin95,gill00,krautter,harman03}, even in the X-rays
\citep[detected a century after outburst by][]{balman}.
The properties of the outflows triggered by nova explosions on massive
WDs are particularly important because they may influence or reveal
the amount of mass lost per explosion, and thereby the ability of the
stars to become type Ia supernovae.


Since the recurrence time for a nova explosion in a given system is
inversely related to the mass of the WD \citep{yaron}, interacting
binaries that contain WDs with masses greater than approximately 1.1
M$_{\odot}$ can have recurrence times of less than 100 years; these
systems are the so-called recurrent novae. In symbiotic recurrent
novae -- of which only 4 are known: \rsoph, T CrB, V3890 Sgr and V745
Sco -- the accreting WD is embedded in a dense nebula fed by the wind
of the red-giant companion.
RS Oph went into outburst for the sixth recorded time (previous
outbursts occurred in 1898, 1933, 1958, 1967, 1985) on February 12th
2006 \citep{narumi06}. The astronomical community observed this
outburst at virtually every wavelength; the X-ray range was extremely
well covered, with observations by $RXTE$ \citep{sokoloski06}, $Swift$
\citep{hachisu07,bode06}, $Chandra$, and $XMM$-Newton
\citep{ness07,nelson08}.

Observations of \rsoph
indicate that the morphology of the expanding shell is complex and
highly asymmetric. Within a few weeks of the start of the 2006
eruption, X-ray spectra showed that the blast-wave evolution deviated
from self-similar, spherical expansion \citep{sokoloski06,bode06}.
Bipolar structure along the east-west direction was clearly seen in
radio images taken a few weeks into the 2006 outburst
\citep{obrien06,rupen08};
several months after the start of the outburst, observations showed
that the radio jets were composed of highly
collimated outflows that powered synchrotron-emitting lobes \citep{jeno08}. 
At optical wavelengths, 
[Ne V]$\lambda$3426 images 5 months after the start of the outburst
revealed an expanding nebular remnant with a double-ring structure
\citep{bode07}. The major axis of the double ring was also oriented
east-west.  Near infrared (NIR) interferometric observations also
revealed extended asymmetric structure in the early expanding shell ---
but along a different direction.
On day 5.5 after the start of the outburst, \citet{chesneau07} found
that the NIR emission arose from a rapidly expanding elliptical
region with major axis oriented from northwest to
southeast. NIR observations in the first month of the outburst by
\citet{lane07} also revealed an asymmetric structure with a position
angle (P.A.) of approximately 120$^{\circ}$/300$^{\circ}$.


In this letter, we describe the first detection of extended X-ray
structure from RS Oph. Taken one and half years after the start of the
2006 eruption, our $Chandra$ observations reveal a narrow X-ray
feature along the direction of the NIR extended emission.
We describe the observation in section \S\ref{sec:obs}, detail the
analysis and results in section \S\ref{sec:image}, and discuss
possible interpretations of the results in section
\S\ref{sec:discuss}.


\section{Observations}
\label{sec:obs}

On 2007 August 4, the $Chandra$ X-ray Observatory (CXO) observed
\rsoph for 90.1 ks using the ACIS-S\ S3 back illuminated chip with a 1/4
subarray (ObsId 7457, start time 06:11:26 UT), in Timed Exposure
mode and
Faint telemetry format. 
The chip was read out every 0.94104 s, with the time between each read
consisting of a 0.9 s exposure and 0.04104 s for charge
transfer. \rsoph produced an average count rate of 0.0682 counts
s$^{-1}$,
or 0.0642 counts per frame integration time. 
The configuration we used
ensured a 
minimal pile-up fraction
(PIMMS\footnote{\url{http://heasarc.nasa.gov/docs/software/tools/pimms.html}}
calculates a negligible pile-up fraction of $2\%$).

\section{Data analysis and results}
\label{sec:image}

\subsection{X-ray spectrum}

With a source count rate of 0.0682 c s$^{-1}$ (see \S~\ref{sec:obs}),
the background subtracted spectrum contained 
6252 counts. The X-ray spectrum of RS Oph in quiescence is complex.
Here we use a preliminary fit ($\chi^2_\nu$=1.35 for 154 d.o.f.) 
consisting of two soft, relatively unabsorbed thermal plasma components ({\tt
  mekal}) and a hard, heavily absorbed cooling flow ({\tt mkcflow})
component \citep[using the model names found in XSPEC;][]{xspec}.  This
fit adequately models the observed spectrum 
for the purposes of the PSF deconvolution processes.  We will present
a full analysis and interpretation of the spectrum in a forthcoming
paper (Nelson et al. in preparation).

\subsection{Extended X-ray structure}
\label{sec:struc}

Since observations at other wavelengths suggested that any extended
structure would be on the order of an arcsec in length,
we applied a Subpixel Event Repositioning \citep[SER;][]{li03,li04}
technique to obtain the best possible spatial resolution. On the basis
of a physical model of the photon--CCD interaction, SER uses the
available information concerning photon event charge splitting to
reconstruct the most likely entrance positions of photons within
individual CCD pixels.  This procedure potentially improves the
effective $Chandra$/ACIS pixel size from 0$^{\prime\prime}$.5 to as
small as 0$^{\prime\prime}$.3, albeit with a non-uniform shape and a
strong dependence on event charge split (i.e., event grade); see
\citet{li03} for details.
Due to the significant number of counts detected, we 
used a spatial bin size of 0$^{\prime\prime}$.125 in constructing the
X-ray images analyzed here, so as to make best use of the potential
improvement in spatial resolution afforded by SER.
After applying SER, we performed a maximum likelihood deconvolution on
the image using a synthetic Point Spread Function (PSF) appropriate
for the source off-axis angle and X-ray spectrum.
Altogether, the process for generating the final image consisted of
the following steps: 1) extract and fit the spectrum
; 2) use the $Chandra$ Ray Tracer \citep[ChaRT;][]{carter03}, with the
spectral model from \#1, to trace the rays through the $Chandra$ X-ray
optics, and then
use MARX (Model of AXAF Response to X-rays) to project those rays onto
the detector and create a synthetic PSF
3) apply SER to the pipeline-processed event file
; 4) use the synthetic PSF from step \#2 and the SER-processed event
file from step \#3 to perform maximum likelihood deconvolution.

\begin{figure*}
 \begin{center}

\includegraphics[scale=0.73]{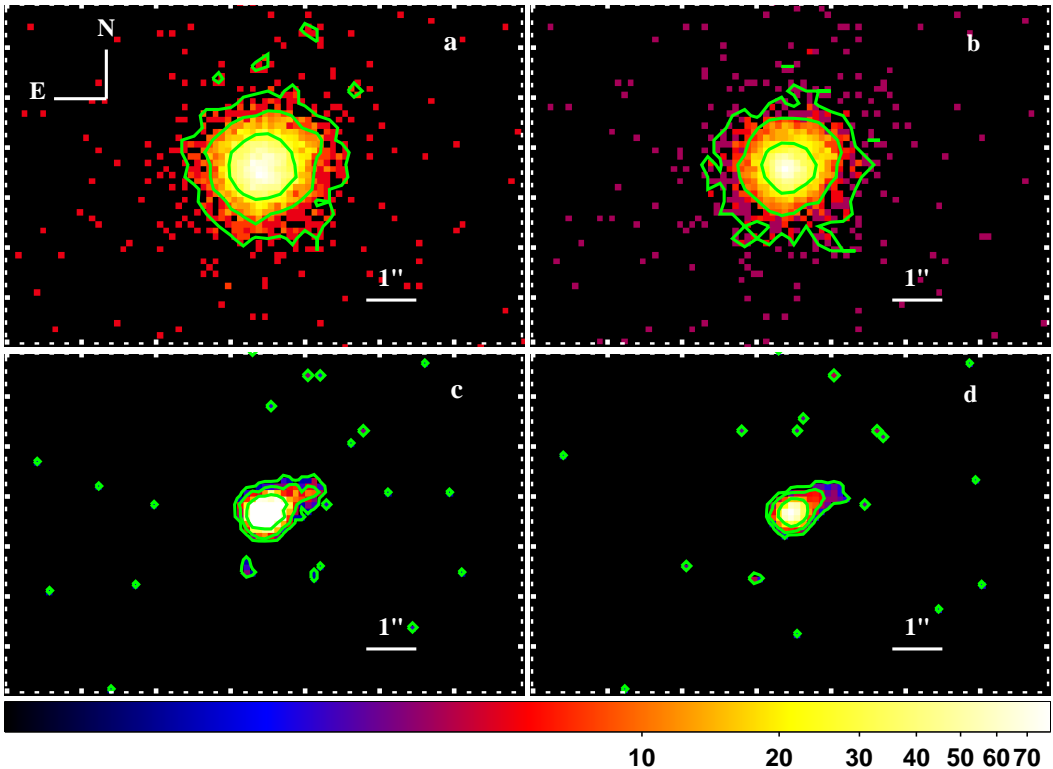}
\caption{Comparison of $Chandra$ images of RS Oph with various levels
  of image processing. (a) Image obtained from pipeline products
  without further processing; (b) Image created after applying
  Subpixel Event Repositioning (SER, see text) to the data; (c)
  Resulting image after performing maximum likelihood deconvolution
  using a simulated PSF; (d) Image created after applying both SER and
  maximum likelihood PSF deconvolution as described in
  \S\ref{sec:struc}.  A comparison of panels (a) and (b) confirms that
  SER does not introduce any spurious asymmetry into the image. A
  comparison of panels (c) and (d) shows that the extended structure
  was clearly visible in the deconvolved image, even before the
  application of SER.  
The innermost contour encloses 4095, 3739, 4955, and 5140 counts in
panels (a) through (d), respectively.  Between the innermost and middle
contours are 1252, 1589, 649, and 489 counts for panels (a) through (d),
respectively.  Between the middle and outermost contours are 299, 326,
66, and 64 counts for panels (a) through (d), respectively.  North is up
and east is to the left.
}
\label{fig1}
 
 \end{center}

\end{figure*}
\begin{figure}
\begin{center}
\includegraphics[scale=0.41]{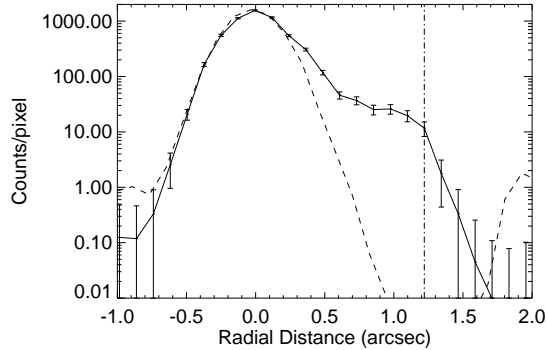}
\caption{Profiles of the intensity along, and perpendicular to, the
  narrow X-ray feature.
The solid curve shows the intensity profile along the narrow feature
while the dotted line shows the profile in the direction perpendicular to
the feature. The length of the extended feature, as measured from the
centroid of the PSF to the location where the intensity falls below 10
counts per 0$^{\prime\prime}$.125 spatial bin (marked with a dot-dashed
vertical line), is 1.2$\pm$0.3 arcsec.
Error bars represent one standard deviation. The error in radial
distance was estimated to be the best achievable SER pixel size, i.e.,
0$^{\prime\prime}$.3.}
\label{fig3}
\end{center}
\end{figure}
\begin{figure}
\begin{center}
\includegraphics[scale=0.53]{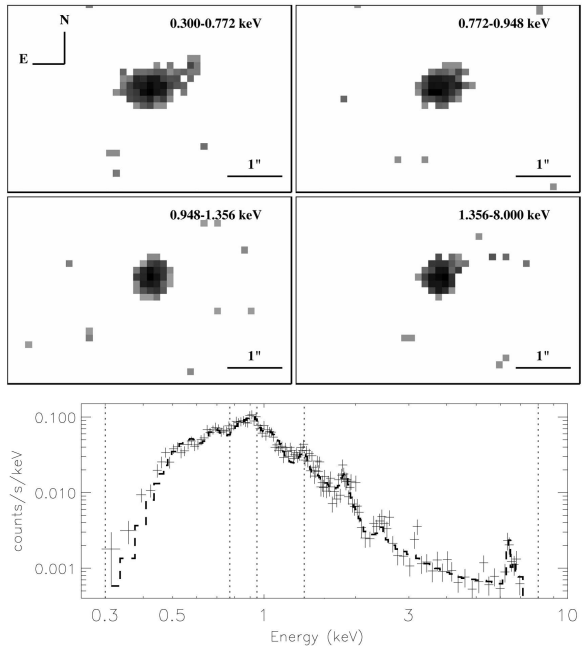}
\caption{Images (to which both SER and maximum likelihood
  deconvolution have been applied) in 4 energy bands containing
  roughly the same number of counts (1000). Lower panel: $Chandra$
  spectrum of RS Oph (and overlaid best-fit model) with dashed
  vertical lines indicating the 4 energy ranges used to generate the 4
  images in the upper panels.
The asymmetric X-ray feature is most pronounced in the soft band ($E <
0.77$ keV). }
\label{fig2}
\end{center}

\end{figure}

The maximum likelihood deconvolution of the $Chandra$ observation
reveals a feature
extending northwest of the central binary ($\alpha$=17h 50m 13.2s,
$\delta$=-06$^{\circ}$ $42'$ $28.2''$) at a P.A. (east of north) of
approximately 300$^{\circ}$ (Fig.~\ref{fig1} and \ref{fig3} ). The
length of the extended feature, as measured from the centroid of the
central X-ray emission to the location where the intensity falls below
10 counts per 0$^{\prime\prime}$.125 pixel, is 1.2$\pm$0.3 arcsec. We
took the uncertainty on the length of the extended structure to be the
best achievable SER pixel size, i.e., 0$^{\prime\prime}$.3.  A
comparison among images from four different energy bands (selected to
contain roughly 1,000 counts each) showed that the narrow X-ray
feature
is primarily composed of photons with energies less than about 0.8 keV
(see Fig.~\ref{fig2}).


\subsection{Tests of the extended X-ray structure}
\label{sec:test}

Although the X-ray feature lies along the readout direction of the
CCD, extensive testing supports our conclusion that it is not an
artifact but is instead representative of the actual
X-ray morphology of RS Oph. 
Any excess charge from a readout artifact would appear as a string of
spurious X-ray events which, for bright sources ($Chandra$ POG v.11),
are not removed by standard grade filtering and fall along rows of
pixels with constant y-coordinate in the CCD pixels array (CHIPY).
Readout streaks are therefore always symmetric and might appear 
along pixels with the same CHIPY value as the source.
Fig.~\ref{fig1} shows that the detected X-ray feature is not symmetric
with respect to the central source.  Furthermore, any readout artifact
in the region of the CCD where we detect the feature would account for
only a few counts, 
whereas we detect
$\sim$ 100 counts in the $\sim1^{\prime\prime}.2$ X-ray feature; we
therefore conclude that it is not due to a readout artifact.

Charge Transfer Inefficiency (CTI) in the S3 chip changes the overall
instrument performance in various ways. First, since some charge is
trapped, the charge read out is less than the amount of charge
deposited.
Second, CTI causes a migration of the grades assigned to each event
due to photons 
that were captured by the traps and gradually {\it
  re-emitted} in the following frames. This second effect can change
grades from that associated with a single pixel events (flight grade =
0) to those associated with a vertical up-split (flight grades = 2,
66), which are oriented parallel to the readout direction (R. Edgars,
private communication). Since no tested CTI-correction algorithm is
available for observations taken in subarray mode, we performed
several tests to indicate the degree to which our observation was
affected by CTI\footnote{The CTI correction applied by the standard
  pipeline process was removed before performing any tests.}.


The S3 back-illuminated (BI) chip has substantial CTI in the serial
array. If severe, this can create an extended feature in the direction
along which the charges move out of the frame store region. In our
observation, such an effect would correspond to a feature aligned in a
direction {\em perpendicular} to the one we detected. On the other
hand, the detected feature could be caused by parallel CTI. To
investigate the presence of parallel CTI, we applied our image
processing procedure to an observation from the $Chandra$ archive that
was taken with the same instrumental configuration (see
\S~\ref{sec:obs}) and has roughly the same count rate as our
observation.  The observation we used was
of HD113703 B (ObsID 626, exposure time = 12.7 ks).
Trapped charges are unlikely to be re-emitted as grade-0 events. 
Since the narrow extended structure from RS Oph appeared strongly when
we created an image using only grade 0 events, we examined an image of
HD113703 B created using only grade 0 events. The grade-0 image of
HD113703 B did not show any extended structure along the read out
direction. This last test suggest that CTI effects were absent.

We also carefully tested whether the image processing that we
performed could have introduced the detected feature as an
artifact. SER is a event-by-event procedure that should not reposition
the events in a preferential direction. Nevertheless, to confirm that
SER did not introduce an artifact,
we applied the deconvolution procedure to data with and without SER
applied (see Fig. \ref{fig1}). No significant differences were found
between the SER and non-SER deconvolved images.
We therefore conclude that the narrow X-ray feature was not a
by-product of applying SER.
If the synthetic PSF was not symmetric, the deconvolution could have
introduced artifacts in the resulting image.  To test for this
possibility, we
applied the deconvolution with various rotations and flips of the PSF
with no significant changes in the orientation of the extended
feature.
Finally, as we discuss below, features at roughly the same P.A. as the
extended X-ray feature have been seen in radio and IR
observations of RS Oph.

\section{Discussion and Conclusion }
\label{sec:discuss}

Assuming, as suggested by our extensive testing, that the extended
X-ray structure is not an artifact, the properties of this feature
indicate that it consists of a few tenths of a percent to a few
percent of the ejecta mass, and that the outermost X-ray emitting
material is moving at a speed approximately half of the escape
velocity from the WD (the escape velocity is
approximately 12,600 km s$^{-1}$
for a 1.35 M$_{\odot}$ WD).
If the material comprising the narrow X-ray feature 
was ejected at the time of the nova explosion, the length of the X-ray
structure suggests an average
expansion rate in the plane of the sky of $\sim$2.3 mas d$^{-1}$. This
rate is similar to the early expansion rate of $\sim$ 2 mas d$^{-1}$
for the radio jet \citep{jeno08}. The rate of expansion of the X-ray
feature corresponds to velocities in the plane of the sky of about
$\sim6,300\pm 1600$~km $s^{-1}$ ($d$/1.6 kpc).
Since the X-ray emitting material is unlikely to be moving exactly in
the plane of the sky, the true velocity could be higher.

A strong shock moving at such a high speed would be expected to
produce hard X-ray emission (with $kT \simeq$ 40 keV in the
strong-shock case); however we find only soft X-ray emission from the
extended feature (\S\ref{sec:struc}).  The discrepancy between the
expected temperature in the strong shock case and the observed photon
energy is also present in the diffuse X-ray emission from planetary
nebulae \citep{soker03} and the two symbiotic stars with X-ray jets --
CH~Cyg \citep{galloway04} and R~Aqr \citep{kellog07}.

The narrow X-ray feature is most extended in the 0.3--0.8 keV energy
band and contains approximately 40 photons in this band.
A crude estimation of its mass can be made if we assume that the
emission is due to an absorbed, optically thin thermal plasma with
$kT$=0.5 keV and n$_{H}$=1.8$\times$10$^{21}$ cm$^{-2}$
\citep{bode06}. We use PIMMS to obtain the observed emission measure
of such a plasma and, assuming that the narrow X-ray feature has a
radius of $\sim$ 0$^{\prime\prime}$.9 and length of
1$^{\prime\prime}$.2 (Fig.~\ref{fig3}), find that the density is
$\sim$40 cm$^{-3}$ and the mass of the extended X-ray emitting
material is $\sim$10$^{-8}$ M$_{\odot}$ (assuming that there is one
proton for every free electron). For comparison, estimates of the
amount of material ejected immediately after the outburst range from
$\sim$10$^{-7}$ M$_{\odot}$ to $\sim$10$^{-6}$ M$_{\odot}$
\citep{sokoloski06,hachisu07}.

\begin{figure*}
\begin{center}
\includegraphics[scale=0.313]{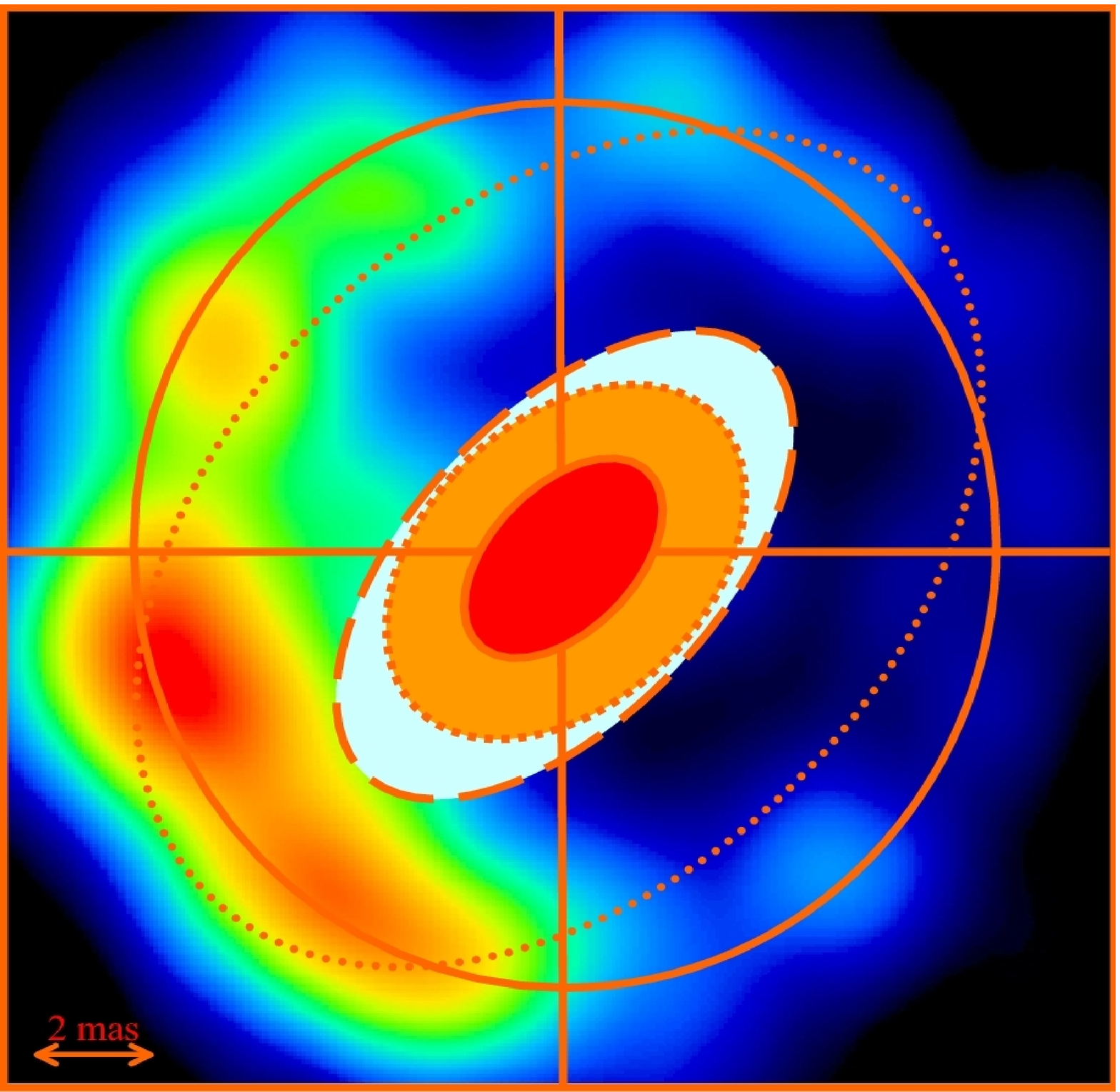}
\includegraphics[height=6.3cm]{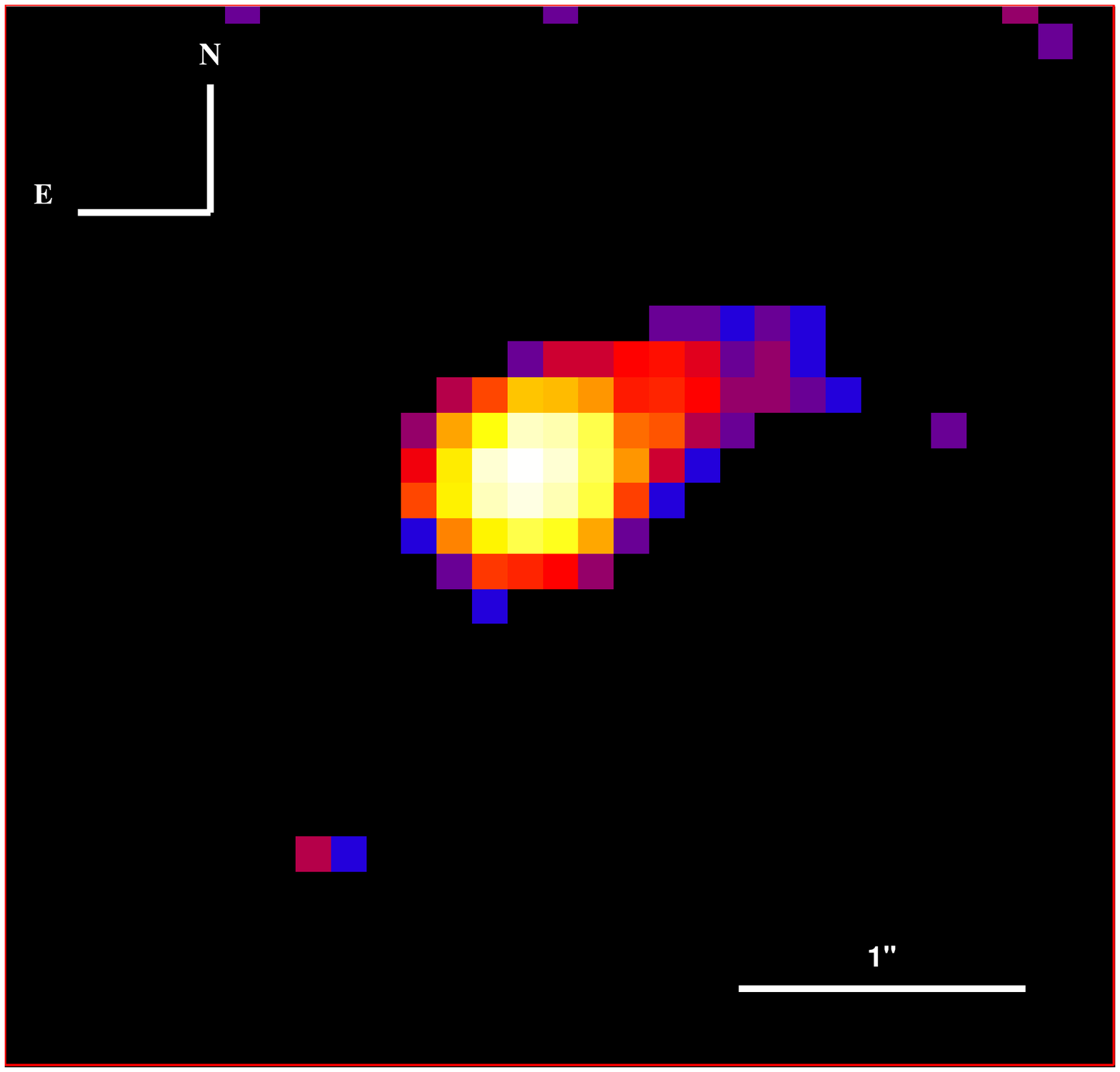}
\caption{Comparison of RS~Oph images in radio, infrared and
  X-rays. {\it Left:} an image of radio synchrotron-emitting plasma
  with NIR contours overlaid \citep{chesneau07}. {\it Right:} the
  X-ray image obtained after SER and PSF-deconvolution processing
  (same as panel (d) in Fig.~\ref{fig1}). The X-ray feature extends
  along the same direction as the NIR and radio features, suggesting
  that they could all be related. Note that the two panels show very
  different scales.}
\label{fig4}
\end{center}
\end{figure*}

The orientation of the narrow X-ray feature is less consistent with
that of the radio jets \citep{obrien06,rupen08,jeno08} than with other
features detected at radio and infrared (NIR) wavelengths.
Early radio observations showed that the ring of synchrotron-emitting
plasma associated with the expanding blast wave had a bright spot at a
P.A.$\approx$120$^{\circ}$ \citep{obrien06,rupen08}. If one were to
superpose the X-ray and radio images and draw a line along the
extended X-ray structure, it would intersect the bright spots on the
radio ring. Moreover, the radio ring was actually more of a C-shape,
with the opening in the ring at the same P.A. as the X-ray feature.
NIR interferometry uncovered extended emission with a P.A. of
120$^{\circ}$/300$^{\circ}$ from day 5.5 to a few months after the
nova explosions \citep{chesneau07, lane07}, and \citet{chesneau07}
suggested a possible connection between the radio and IR morphologies.
Our detection of an X-ray feature extending in the northwest
direction, like features in the radio and NIR, suggests that the
structures along this direction could all be related (see
Fig.~\ref{fig4}).

Although the X-ray structure has some similarities to features
predicted by models of the explosion, many properties of the observed
narrow X-ray feature are unexplained, and the true origin of this
extended X-ray emitting gas remains unclear.  Computational modeling
suggests that the blast-wave expansion should be strongly asymmetric.
For example, in the 3-dimensional simulations of \citet{walder08}, the
blastwave moves into a highly inhomogeneous pre-outburst density
distribution and forms an elongated, bipolar structure perpendicular
to the plane of the orbit.  We do not see any extended X-ray emission
in this direction.
In contrast, models by \citet{orlando09} predict a one-sided X-ray
structure -- as we have found -- due to the reflection of the
blastwave off of the red giant.  
Unfortunately, however, the position of the red giant at the time of
the nova explosion \citep[according to the ephemeris of][]{brandi09}
would have led to an X-ray extension in the north-south direction,
which is inconsistent with our detection of extended X-ray emission at
a P.A. of ~300 degrees.
Thus, our results suggest that either the orbital solution for this
system needs to be re-examined or that the explosion somehow ejected
X-ray emitting plasma in a direction that was neither perpendicular to
the plane of the orbit nor along the line joining the WD and the red
giant.

\acknowledgments We thank L. Townsley, F. Bauer, E. Gotthelf,
R. Edgars, and M. McCollough for discussions about the ACIS CCDs,
T. Nelson for discussion about spectral results, and O. Chesneau for
providing us a good quality version of Fig.~4b. Support for this work
was provided by NASA through Chandra awards GO7-8030X and GO6-7022A
issued by the Chandra X-ray Observatory Center, which is operated by
the SAO for and on behalf of NASA under contract NAS8-03060.



 {\it Facilities:} \facility{CXO (ASIS)}.

\end{document}